# Quantum Metrology: Towards an alternative definition for the meter

Ioannis P. ZOIS

Testing, Research & Standards Centre, Public Power Corporation, 9, Leontariou Street, GR- 153 51, Pallini, Athens, Greece

i.zois@dei.com.gr and i.zois@exeter.oxon.org

**Abstract**. The motivation for this work came from an attempt to give an alternative definition for the meter, the SI unit for measuring length. As a starting point towards this goal we present the underlying theory behind our approach which uses ideas from quantum field theory and non-commutative geometry, in particular the notion of an odd K-cycle which is based on the Dirac operator (and its inverse, the Dirac propagator). Using physics terminology, the key point in our strategy is this: instead of measuring ordinary length in space-time we measure the "algebraic (or spectral) length" in the space of quantum states of some fermion acted upon by the Dirac propagator.

## 1. Introduction

The SI unit for measuring lengths is the meter. This is defined as the length of the path traveled by light in vacuum during a time interval of 1/299,792,458 second (see [3], [25], [26], [17]). It is clear that this definition is based on A. Einstein's (Special and General) Theory of Relativity (see [6], [10], [23], [27], [28], [29]). A basic ingredient of the theory of relativity as much as physics is concerned is the fact that the speed of light is constant, namely it is the same for all inertial observers (this follows from numerous experiments, starting with the famous Michelson - Morley experiment in 1887). In fact the fundamental difference between the relativity principle of Einstein and that of Galileo is the exchange of the invariance of simultaneity with the invariance of the speed of light.

The mathematics used in relativity theory is (pseudo) - Riemannian Geometry (see [9], [15], [21]). The mathematical structure studied in Riemannian geometry is the Riemannian manifold (or Riemannian space): A Riemannian manifold is a pair *(M, g)* where     is a real smooth manifold whose tangent space at each point is equipped with a Riemannian metric *g,* that is a positive definite symmetric bilinear form (inner product) which varies smoothly on M. If g is not positive definite then one has a pseudo-Riemannian metric.

A metric enables one to measure lengths, angles, areas, volumes, to calculate curvatures, gradients of functions, divergences of vector fields etc. The space-time we live in is an example of a pseudo-Riemannian manifold of dimension 4. Since by definition a manifold locally looks like the Euclidean space $\mathbf{R}^n$ and since the dimension of space-time is 4 (at least macroscopically), then our space-time locally looks like $\mathbf{R}^4$. An important geometric difference between the Special (abbreviated to SR) and the General (abbreviated to GR) Theory of Relativity is the fact that in the former space-time is flat, namely both locally and globally is modeled on $\mathbf{R}^4$ whereas in the later space-time is a pseudo-Riemannian manifold with some nonzero curvature and hence it is only locally modeled on $\mathbf{R}^4$.

During the last 20 years or so an important break through occurred in geometry, this is the discovery of noncommutative geometry (see [4], [5], [31], [32], [33]). This new geometry enormously generalizes Riemannian geometry and it has deep connections to quantum physics (quantum field theory in particular). These new (as far as metrology is concerned) ingredients (noncommutative geometry and quantum field theory) constitute the basis of our work: Our idea is we believe quite clear, we imitate what happens in *cartography:* Instead of measuring the distance between say London and Paris, we use a map with its scale S. We measure the distance on the map DM and then we can calculate the real distance RD by the simple formula

$$RD = DM \ X \ S.$$

In the same fashion, instead of measuring the real spacetime distance (distance on a Riemannian manifold), we use an equivalent "algebraic (or spectral) distance" on a mathematical structure called "odd K-cycle of the corresponding C*-algebra" of the Riemannian manifold. This plays the role of the map with scale. A key ingredient of this structure is the Dirac operator used in quantum physics to describe fermions (relativistic Dirac equation). Then the proposition proved serves as a "translator" converting distances (and data) from the real spacetime (Riemannian manifold) to this fictitious space (odd K-cycle). It's the analogue of the simple equation used in cartography.

There is another motivation for our approach though: During the *24$^{th}$ General Conference of Standards and Measures on 21$^{st}$ October 2011 in Serves France,* there was an unanimous vote for the redefinition of several of the fundamental units, some of them based on the Planck's constant h. It is common knowledge nowadays that Planck's constant is the fundamental constant in quantum physics, it captures the so called particle -wave duality, the fundamental principle of quantum physics. Mathematically speaking Planck's constant measures the noncommutativity between position and momentum as expressed in Heisenberg's uncertainty principle, thus the connection with noncommutative geometry. One could say that this is the official introduction of Quantum Physics in metrology. Our approach relies heavily on quantum field theory.

The purpose of this article is to set the mathematical (geometric) basis for a possible application of this new geometry and quantum physics to metrology (hence the term quantum metrology) and especially to the definition of the fundamental units. We hope this will be the beginning of a fruitful interaction between mathematical physics and metrology.

## 2. The underlying geometry of the current definition of the meter
*2.1 Minkowski geometry*

In SR space-time is modeled on the vector space (but also trivially a manifold) $\mathbf{R}^4$ equipped with the so-called Minkwski (pseudo) metric. The fact that one needs the Minkowski and not the usual Euclidean metric in 4 dimensions can be readily seen for example from the wave equation that the electromagnetic field satisfies in vacuum:

$$\Box \ _\mu = 0, \qquad (1)$$

where $\Box$ is the d'Alembert operator:

$$\Box = (1/c^2)dt^2 - \nabla^2 \quad (2)$$

and $A_\mu$ is the 4-potential. d'Alembert operator is nothing other than the Laplace operator in $\mathbf{R}^4$ using the Minkowski metric. It is well-known that one of the theoretic motivations for the discovery of SR is the fact that Maxwell's equations are not compatible with Galileo's relativity principle.

Choosing a basis, symmetric bilinear forms can be represented by symmetric matrices, in our case by a real 4 × 4 matrix. If we pick a $\eta$-Cartesian basis (an inertial reference frame as it is known in physics literature), Minkowski metric can be represented in its canonical form by the matrix

$$\eta = \mathrm{diag}\,(1, -1, -1, -1) \quad (3)$$

where we adopt the signature (+ - - -). Hence the inner product between two vectors with coordinates (relative to the chosen basis) $x = (x^0, x^1, x^2, x^3)$ and $y = (y^0, y^1, y^2, y^3)$ will be given by the following expression

$$\eta(x, y) = x^0 y^0 - x^1 y^1 - x^2 y^2 - x^3 y^3 \quad (4).$$

Given a metric it is clear that one can define the length of a vector $x = (x^0, x^1, x^2, x^3)$ (in physics terminology since these are elements of $\mathbf{R}^4$ they are called 4-vectors), which will be equal to

$$|x| = \eta(x, x) = x^0 x^0 - x^1 x^1 - x^2 x^2 - x^3 x^3 = (x^0)^2 - (x^1)^2 - (x^2)^2 - (x^3)^2 \quad (5)$$

One can immediately define angles, for example the cosine of an angle say $\theta$, between two vectors $x$ and $y$ will be given by the relation

$$\cos\theta = \eta(x, y) / (|x| \, |y|) \quad (6)$$

In physics one uses the (infinitesimal) line element (proper-time)

$$d\tau^2 = (c\,dt)^2 - dx^2 - dy^2 - dz^2 \quad (7)$$

Since the Minkowski metric is not positive definite it is apparent that expressions like (4), (5) and (7) can take positive, negative and zero values. In physics one uses the terms time-like, space-like and null respectively to characterize 4-vectors. For reasons which are out of the scope of this article, in SR we accept that only time-like vectors have physical meaning (namely only time-like vectors can be used to describe physical quantities) and that null 4-vectors describe photons (particles traveling with the speed of light). Thus if one wants to describe the path (position vector) of a photon, equation (7) will give

$$0 = (c\,dt)^2 - dx^2 - dy^2 - dz^2$$

hence

$$(c\,dt)^2 = dx^2 + dy^2 + dz^2$$

thus

$$(c\,dt)^2 = ds^2 \quad (8)$$

Equation (8) gives the current definition of the meter in SI stated at the start of section 1 (this is the infinitesimal version of the equation, the finite version is obtained easily by integration).

*2.2 Riemannian geometry*

In the case of Riemannian geometry things are generalized as follows: To begin with, instead of the space $\mathbf{R}^4$ one has a real smooth manifold, we briefly recall its definition: A real smooth manifold of dimension (in general) n is a Hausdorff topological space which is locally homeomorphic to the Euclidean space $\mathbf{R}^n$, that is for every point of the manifold there exists a neighborhood (an open set containing it) which is homeomorphic to an open ball of dimension n (homeomorphic means that there exists a map which is 1-1, onto, continuous with a continuous inverse). These maps which locally model the manifold onto the space $\mathbf{R}^n$ are called coordinate charts. It is clear that coordinate charts may intersect and any point on the manifold can be arranged to belong to different charts. In smooth manifolds the extra condition is that the maps which represent these coordinate transformations on intersections should be smooth (since these coordinate transformations are essentially maps from $\mathbf{R}^n$ to $\mathbf{R}^n$, the term "smooth" means the usual thing we know from vector calculus, that is infinitely times differentiable). This definition enables one to carry over the usual vector calculus onto manifolds which constitute an important generalisation of spaces like $\mathbf{R}^n$. The classic example of a smooth real manifold of dimension two is the 2-sphere $S^2$.

The smoothness condition guarantees that the manifold has a well-defined tangent space at each point (this is a simple higher dimensional generalisation of the tangent vector of a smooth curve). The tangent space is nothing other than a vector space $\mathbf{R}^n$. Hence a Riemannian metric is a positive definite symmetric bilinear form on the tangent space which varies smoothly from point to point. Consequently our discussion in the previous section concerning metrics (symmetric bilinear forms) on vector spaces $\mathbf{R}^n$ can be immediately applied to this more general setting. Hence a Riemannian manifold is a pair (M,g) where     is a smooth real manifold and g is a Riemannian metric. It is clear that in SR the space-time manifold can be identified with its tangent space whereas in GR this is not necessarily true (compare $\mathbf{R}^2$ with the 2-sphere $S^2$, $\mathbf{R}^2$ is the tangent space on every point of the 2-sphere).

If one chooses a local coordinate system on a Riemannian manifold (the analogue of a basis in a vector space), then the metric can be represented by a symmetric n X n matrix yet the entries of this matrix will no longer be numbers but functions of the local coordinates (since the metric varies from point to point). More concretely, according to tensor algebra, the metric is a covariant tensor (field) of order 2. Then the line element on a Riemannian manifold (  , g) of dimension n, by choosing local coordinates denoted $x^\mu$ (where $\mu,\nu = 1,2,..,n$), will be given by the following formula
$$ds^2 = g_{\mu\nu} dx^\mu dx^\nu \quad (9)$$
where we have made use of the Einstein convention (repeated indices are summed). It is a fundamental physical principle that GR locally reduces to SR (this fact makes the notion of manifold appropriate to describe space-time itself, at least classically and at long scales). Furthermore photons (namely light) according to GR follow geodesics, in Riemannian geometry geodesics are the analogue of straight lines in Euclidean geometry, these are the paths - curves of minimal (more accurately extreme) length. Thus the geodesic distance d(P,Q) between two points P, Q on some Riemannian manifold is given by the length of the geodesic which connects the two points using equation (9). Essentially it is the infimum of the length of all possible paths connecting the two points P and Q where the length of each path is calculated using equation (9).

## 3. Noncommutative geometry

Noncommutative geometry is perhaps the most important development in geometry during the last 20 years or so. It was discovered by Alain Connes and Daniel Quillen.

The starting point of noncommutative geometry (abbreviated to NCG in the sequel) is Gelfand's theorem which we briefly explain: One the one hand let's consider the category whose objects are compact Hausdorff topological spaces and whose morphisms (arrows) are homeomorphisms among them. On the other hand let's consider the category whose objects are unital commutative C*-algebras and whose morphisms (arrows) are star preserving homomorphisms. Gelfand's theorem then states that these two categories are equivalent and the equivalences are given by the (mutually inverse) functors C(-) and Spec(-). The former assigns to each compact Hausdorff topological space the commutative unital C*-algebra C(X) consisting of continuous complex valued functions defined on X. The later assigns to every commutative unital C*-algebra the compact Hausdorff topological space Spec(A) which is the spectrum of the algebra A (the spectrum of an algebra consists of its characters, namely non-zero star preserving homomorphisms from the algebra to the field of the complex numbers **C).** These functors satisfy the following conditions

$$\text{Spec} \circ C = 1$$

and

$$C \circ \text{Spec} = 1,$$

namely C(Spec (A)) = A and Spec (C(X)) = X, where 1 is the identity functor and " ∘ " denotes the composition of functors. If the topological space is not compact but only locally compact, then the corresponding algebra has no unit (this difficulty can be overcome by adjoining a unit which increases complexity).

Gelfand's theorem implies that the topological information of a space is encoded in the corresponding algebra of continuous complex functions and in fact the topology can be recovered from the algebra. In particular this holds true for the topology of a Riemannian manifold. The natural question then is if there is a way to recover the metric data as well. It turns out that this can indeed be done using the notion of an (odd) K-cycle (see [22]). The definition is the following:

*Definition 1:* An (odd) -cycle on an algebra A is a quadruple (A, , H, D) where is an algebra (commutative or not), is a separable Hilbert space, is a unitary representation of the algebra A on the Hilbert space and D is an unbounded self-adjoint operator acting on the Hilbert space H satisfying the following properties:

1. The commutator [D,a] = Da - aD (where a is an element of the algebra ) is a bounded operator for any element a of the algebra A.
2. The operator $(1+D^2)^{-1}$ is compact.

(If the algebra A is non-unital then we slightly modify the above definition by demanding that the operator $[a(1+D^2)]^{-1}$ is compact for any element a of the algebra A).

The way one can recover the metric data using the notion of a K-cycle is deferred to section 5 where it will be presented as a Lemma.

K-cycles are the basic ingredients of K-Homology (see [31], [16]). This theory is the dual theory to the topological K-Theory due to Atiyah and Grothendieck (see [1], [11] and [13]) which has been proved a very useful tool in topology, geometry but also in modern theoretical physics (superstring theory, M- Theory, the theory of branes etc).

**4. The Dirac operator**

The Dirac operator will play a central role in our discussion and hence it deserves a brief review. Elementary particles belong to two big families, the bosons and the fermions. According to the spin statistics theorem in theoretical physics (see [24], we are not interested here in anyons, braid statistics etc), the former have integer spin and the obey the Bose-Einstein statistics whereas the later have half-integer spin, the obey the Dirac-Fermi statistics and the Pauli exclusive principle. Bosons are described by the Klein-Gordon equation (a second order partial differential equation of hyperbolic type) which is the relativistic version of the Schrodinger equation used in quantum mechanics (essentially this is the Laplace operator in 4-dimensional Minkowski space with signature (+ - - -) ):

$$(\Box + \mu^2)\psi = [(1/c^2)\partial_t^2 - (\partial_x^2 + \partial_y^2 + \partial_z^2) + \mu^2]\psi = 0 \quad (10)$$

where $(\partial_x^2 + \partial_y^2 + \partial_z^2)$ is the Laplace operator in 3 spatial (Euclidean) dimensions (using Cartesian coordinates) and $\mu = 2\pi mc/h$ (m denotes the rest mass, c denotes the speed of light and h denotes Planck's constant).

Fermions again in 4-dimensional Minkowski space are described by Dirac's equation (this is written in the original Dirac's form):

$$(\beta mc^2 + a_k P_k c)\psi = [(ih)/(2\pi)]\psi_t \quad (11)$$

The index k takes the values 1,2,3 (these are the three spatial coordinates), m denotes the rest mass of the fermion, c denotes the speed of light and P denotes the momentum in the quantum mechanical sense, namely $P_k$ corresponds to the partial derivative with respect to the corresponding spatial coordinate

$$P_k \to -ih\partial_k.$$

The "mysterious" quantities $\beta$ and $a_k$ are in fact complex $4 \times 4$ matrices satisfying the relations

$$(a_k)^2 = \beta^2 = I_4 \quad (12)$$

where $I_4$ denotes the identity $4 \times 4$ matrix and

$$a_i a_j + a_j a_i = 0 = a_i \beta + \beta a_i \quad (13)$$

(all indices i,j,k take values 1,2,3). One can rewrite Dirac's equation in covariant relativistic form as follows (see [7], [8], [20], [14] and [30]):

$$[(-ih/2\pi)\gamma^\mu \partial_\mu + mc]\psi = 0 \quad (14)$$

where $\mu = 0,1,2,3$ and $\gamma^\mu$ denote the famous Dirac matrices (these are complex $4 \times 4$ matrices) which constitute a basis for the 4-dimensional Clifford algebra (in particular this is called the Dirac algebra) and they satisfy the relations

$$\{\gamma^\mu, \gamma^\nu\} = 2\eta^{\mu\nu} \quad (15)$$

The symbol {,} denotes the anti-commutator (namely $\{\alpha, \beta\} = \alpha\beta + \beta\alpha$) and $\eta^{\mu\nu}$ denotes the (contravariant) Minkowski metric we saw in section 2. The following relation holds

$$\gamma^\mu \gamma_\mu = \delta_\mu^\mu \quad (16)$$

where $\delta_\mu$ denotes the Kronecker $\delta$.

In order to relate the covariant form with Dirac's original form of the equation we set $\beta = \gamma^0$ and $\gamma^k = \gamma^0 a^k$ (indices are lowered and raised using the metric):

$$\gamma^0 = \begin{pmatrix} I_2 & 0 \\ 0 & -I_2 \end{pmatrix},$$

$$\gamma^1 = \begin{pmatrix} 0 & \sigma_x \\ -\sigma_x & 0 \end{pmatrix},$$

$$\gamma^2 = \begin{pmatrix} 0 & \sigma_y \\ -\sigma_y & 0 \end{pmatrix} \text{ and}$$

$$\gamma^3 = \begin{pmatrix} 0 & \sigma_z \\ -\sigma_z & 0 \end{pmatrix}$$

where again $I_2$ denotes the identity 2 × 2 matrix and $\sigma_x$, $\sigma_y$, $\sigma_z$ denote the Pauli 2 × 2 complex matrices.

The quantity $\gamma^\mu \partial_\mu$ is called the *Dirac operator*. The Dirac equation gave a precise description of the electron (relativistic particle of spin ½), it explained accurately the emission spectrum of the hydrogen atom (especially the hyperfinite structure) and it predicted the existence of antimatter. From the mathematical viewpoint the Dirac equation can be explained as an eigenvalue equation where the rest mass is proportional to the eigenvalue of the momentum operator. The Dirac operator can be interpreted as the square root of the Laplace operator. Let us also note here that all mass particles known today (quarks, leptons and neutrinos) are fermions and hence they are described by the Dirac equation.

The inverse of the Dirac operator is called the *Dirac propagator* in physics (or 2 point function). This is nothing other than the Green function for the (linear) Dirac operator. In position space this is given by the following formula

$$S_F(x-y) = \int \frac{d^4p}{(2\pi)^4} e^{-ip(x-y)} \frac{\gamma^\mu p_\mu + m}{p^2 - m^2 + i\varepsilon} = \left(\frac{\gamma^\mu (x-y)_\mu}{|x-y|^5} + \frac{m}{|x-y|^3}\right) J_1(m|x-y|) \quad (17)$$

where $J_1(m|x-y|)$ is the Bessel function of 1st kind. By a Fourier transform, in momentum space the Dirac propagator is given by the equation

$$SF(p) = \frac{\gamma^\mu p_\mu + m}{p^2 - m^2 + i\varepsilon} \quad (18)$$

The physical meaning of the propagator is that following Feynman's path integral approach, it is the basic ingredient in the expression that calculates the probability amplitude for a particle (in our case for a fermion) to travel from one place to another in a given time (position space) or to travel with certain energy and momentum (momentum space).

If we have an arbitrary Riemannian manifold (assumed closed for simplicity) instead of 4-dimensional Minkowski space, the above construction can be generalized (in a not so straightforward way). This was done by Atiyah - Hirzebruch - Patodi - Saphiro - Singer (see [19], [12] and [2]) as follows: Given a Riemannian manifold say $M$ of dimension n, the Riemannian metric reduces the structure group GL(n) (invertible n X n real matrices) of the tangent bundle $TM$ of M to the special orthogonal group SO(n) (this is the group of orthogonal n X n real matrices with unit determinant). In simple words this is a mathematical expression of the fact that given a metric one can define lengths and angles and hence one can define orthonormal bases, namely bases whose vectors are mutually vertical with length 1. Next one defines the spin bundle S over $M$ which is the 2:1 lift of the bundle of orthonormal frames. The spin bundle S is a principal bundle with structure group Spin(n), which is the double cover of the group SO(n). If the above lift is possible, then one says that M is a spin manifold. A manifold is spin if the second Stiefel-Whitney class of its tangent bundle vanishes (the manifold is orientable if its 1st Stiefel-Whitney class vanishes).

In order to describe the Dirac operator for Riemannian manifolds one needs to define Clifford modules. More concretely, let (V,Q) be a vector space (over a commutative field k) equipped with a quadratic form Q. The Clifford algebra Cl(V,Q) is an associative algebra with unit which is defined as the quotient of the free tensor algebra $FT(V) = \otimes^r V$ by the ideal I(V) of FT(V) generated by the elements of the form $v \otimes v + Q(v)1$ where v belongs in V and $\otimes$ denotes tensor product, namely:

$$Cl(V,Q) = \otimes^r V / I(V) \quad (19).$$

Let us recall from the theory of Lie groups that for $n \geq 3$ there exists the universal covering homomorphism $\rho_0 : \text{Spin}(n) \to \text{SO}(n)$ with kernel $\{1,-1\}$ (this is the meaning of the statement that the Lie group Spin(n) is the double cover of the Lie group SO(n)). Let us assume that $n \geq 3$ and let $\varepsilon$ be an oriented Riemann vector bundle of dimension n over some manifold . A spin structure on $\varepsilon$ is a principal fibre bundle P(Spin(n)),E with structure Lie group Spin(n) along with a double cover map

$$\rho : P(\text{Spin}(n)),E \to P(\text{SO}(n)),E \quad (20)$$

where P(SO(n),E) denotes the bundle of oriented orthonormal frames of the vector bundle $\varepsilon$ that satisfy the relation $\rho(pg) = \rho(p)\rho_0(g)$ for every element p in P(Spin(n)),E and g an element of the Lie group Spin(n).

A spin manifold is an oriented Riemannian manifold with a spin structure on its tangent bundle (the dimension of the tangent bundle equals the dimension of the manifold).

Now every orthogonal transformation of $\mathbf{R^n}$ induces an orthogonal transformation on the Clifford algebra Cl($\mathbf{R^n}$) which maps the tensor algebra to itself and preserves the ideal. This induced map on the Clifford algebra is a homomorphism (it preserves multiplication) and hence one obtains a representation

$$cl(p(n)): \text{SO}(n) \to \text{Aut}(Cl(\mathbf{R^n})) \quad (21).$$

If E is an oriented Riemann vector bundle as above over some manifold , then the Clifford bundle Cl(E) of E is defined as

$$Cl(E) = P(\text{SO}(n),E) \times_{cl(p(n))} Cl(\mathbf{R^n})$$

where

$$cl(p(n)): \text{SO}(n) \to \text{Aut}(Cl(\mathbf{R^n}))$$

is the representation coming from the induced transformations on the orthogonal transformations on $\mathbf{R^n}$. Since the vector bundle E is Riemann, that means that one has an inner product defined on its fibres and Cl(E) is the associated bundle of Clifford algebras. Alternatively, Cl(E) could be defined as the quotient of the free tensor bundle $\otimes^r \varepsilon$ of $\varepsilon$ by the ideal bundle I($\varepsilon$), where I($\varepsilon$) is the bundle over M whose fibres are the two-sided ideals of the free tensor algebra of the fibre of E generated by the elements of the form

$$v \otimes v + ||v||^2$$

where v is an element of the fibre of E. Hence one has (in analogy with equation (19) for vector spaces) that

$$Cl(E) = \otimes^r \varepsilon / I(E).$$

It is clear that Cl(E) is a Clifford algebra bundle over M and the fibre-wise multiplication on Cl(E) gives an algebra structure on the space of sections $\Gamma$(Cl(E)) of Cl(E).

Let M be a Riemannian manifold of dimension n with Clifford bundle Cl(M) (namely Cl(M) is the Clifford bundle of the tangent bundle TM of M, i.e. Cl(M) = Cl(TM)) and let S denote an arbitrary bundle of left modules over Cl(M) (that is a bundle over M whose fibre $S_x$ over a point x of is a left module over the Clifford algebra $Cl(M)_x$, S is called the spinor bundle). Assume further that S is Riemann with a Riemann connection (connection compatible with the metric on the manifold with vanishing torsion). Under these assumptions one can define in a natural way a first order differential operator D acting on sections of S which is called a *Dirac (type) operator* as follows

$$D = e_j \cdot \nabla_{e_j} \quad (22)$$

where the $e_i,...,e_n$ give an orthonormal basis of $T_xM$, is a section of S, $\nabla$ denotes the covariant derivative on S defined by the connection and "·" denotes the Clifford multiplication. The square $D^2$ of the Dirac operator is called *the (corresponding) Dirac Laplacian*.

## 5. The basic Lemma

The central idea of this article is given in the following Lemma:

*Fundamental Lemma:*
"One can recover the Riemannian metric from the fundamental K-cycle defined by the Dirac operator on a spin manifold ; In particular one has that

$$d(p,q) = \sup \{ | f(p) - f(q) | \, ; f \in C(M) \text{ with } \| [D,f] \| \leq 1 \} \quad (23)$$

where p and q are points in M and d(p, q) denotes their geodesic distance ".

*Proof:*
Let (M,g) be a closed n-dim Riemannian spin manifold. The fundamental K-homology class (A, , , D) of M (an odd K-cycle, see definition in section 3) is defined as follows: As algebra we take the algebra C(M) of complex continuous functions on M (we could take smooth functions as well). Clearly C(M) is a vector space and it can be turned into an algebra using point-wise multiplication:

$$(fg)(x) = f(x)g(x) \quad (24)$$

where f, g are elements of C(M) and x is a point of (note that in the LHS of (24) we have multiplication in C(M) whereas in the RHS of (24) we have the usual multiplication of complex numbers). In fact C(M) is a unital C*-algebra where the unit is given by the constant function 1, star "*" operation is just complex conjugation and the norm $\|.\|$ is given by

$$\|f\| = \max \{ |f(x)| \, \forall x \in M \}. \quad (25)$$

It is easy to check that C(M) is complete with respect to the metric induced by the above norm (well, in fact this is part of Gelfand's theorem).
As Hilbert space H we take the space $H = L^2(M,S)$, namely the space of square integrable sections of the (irreducible) spinor bundle S of M, its dimension is equal to $2^{(n/2)}$. The scalar product in $H = L^2(M,S)$ is the usual one of the measure $d\mu(g)$ associated with the Riemannian metric g, i.e.

$$( , ) = \int d\mu(g) \, ^*(x) \, (x) \quad (26)$$

where , are elements of the Hilbert space H, "*" denotes complex conjugation and the scalar product in the spinor space is the natural one in $\mathbf{R}^k$ where $k = 2^{(n/2)}$.

As unitary representation we take the action of C(M) as multiplicative operators on H: If f is an element of C(M) and an element of , then $(f\,)(x) = f(x)\,(x)$.

Finally, as self-adjoint operator D we take the Dirac operator D of the spin manifold M associated with the Levi-Civita connection $= dx^\mu \,_\mu$ of the metric g. It is worth describing the Dirac operator in greater detail using local coordinates: Let ($e_a$, a = 1,2,...,n) be an orthonormal basis of vector fields related to the natural basis ($\partial_\mu$, $\mu = 1,2,...,n$) via the n-beins $e_a^\mu$ so that the components of the metric g are related by
$$g^{\mu\nu} = e_a^\mu e_b^\nu \eta^{ab} \quad (27)$$
where $\eta^{ab}$ denotes the flat metric (the Greek indices $\mu,\nu$ are the "curved" indices and the Latin indices a,b are the "flat" indices, all take values $1,2,..,n = \dim M$, curved indices are lowered and raised using the Riemann metric g whereas flat indices are lowered and raised using the flat metric $\eta$). The coefficients ($\omega_{\mu a}^{\ b}$) of the Levi-Civita connection (metric and torsion free) of the metric g are defined via
$$\nabla^\mu e_a = \omega_{\mu a}^{\ b} e_b \quad (28)$$
Moreover let Cl(M) denote the Clifford bundle over M whose fibre over x in M is the (complexified) Clifford algebra Cl($T_xM$) and let us denote by $\Gamma(\ ,Cl(M))$ the module of its sections. One gets an algebra homomorphism
$$\gamma : \Gamma(\ , Cl(M)) \to \mathcal{L}(\mathcal{H})$$
defined by
$$\gamma(dx^\mu) = \gamma^\mu(x) = \gamma^a e_a^\mu \quad (29)$$
The curved gamma matrices satisfy a similar relation with equation (15):
$$\{\gamma^\mu(x), \gamma^\nu(x)\} = 2 g^{\mu\nu} \quad (30)$$

The lift $\nabla^s$ of the Levi-Civita connection to the spinor bundle will be
$$\nabla^s_\mu = \partial_\mu + \omega_\mu^s = \partial_\mu + (1/2) \omega_\mu^{ab} \gamma_a \gamma_b \quad (31)$$
Then the Dirac operator defined by
$$D = \gamma \nabla \quad (32)$$
can be written in local coordinates as
$$D = \gamma(dx^\mu) \nabla^s_\mu = \gamma^\mu(x)(\partial_\mu + \omega_\mu^s) = \gamma^a e_a^\mu (\partial_\mu + \omega_\mu^s) \quad (33)$$

It is straightforward to check that the conditions of the definition of an (odd) K-cycle are satisfied and thus (C(M), $L^2$(M,S), $\cdot$, D) is an (odd) K-cycle (this is an odd class in the K-Homology of M, namely an element of the 1$^{st}$ K-homology group $K_1(\ )$ of ).

In order to recover the geodesic distance one works as follows: From the action of C(M) as multiplicative operators on $H = L^2$(M,S) one finds that
$$[D, f]\psi = (\gamma^\mu \partial_\mu f)\psi$$
where $\psi$ is in the Hilbert space and f is in C(M). Moreover the commutator [D, f] is itself a multiplicative operator as well, namely
$$[D, f] = (\gamma^\mu \partial_\mu f) = \gamma(df)$$
As a consequence its norm is
$$\| [D,f] \| = \sup |(\gamma^\mu \partial_\mu f)(\gamma^\nu \partial_\nu f)^*|^{1/2} = \sup |\gamma^\mu(\partial_\mu f)(\partial_\nu f)^*|^{1/2} \quad (34)$$
Now the RHS of equation (34) is just the Lipschitz norm of f (see [4]) which is given by

$$\|f\|_{Lip} = \sup_{x \neq y} \frac{|f(x)-f(y)|}{d_\gamma(x,y)} \quad (35)$$

where $d_\gamma(x,y)$ is the usual geodesic distance on M given by the well-known formula
$$d_\gamma(x,y) = \inf \{\text{length of paths } \gamma \text{ from x to y}\}.$$
Hence one has that

$$\|[D,f]\| = \sup_{x \neq y} \frac{|f(x)-f(y)|}{d_\gamma(x,y)}.$$

Now the condition $\|[D,f]\| \leq 1$ in the Lemma (equation (23)) automatically gives that
$$d(p,q) \leq d_\gamma(p,q) \quad (36)$$
To invert the inequality sign we fix a point q and we consider the function
$f_{,q}(x) = d(x,q)$. Then $\|[D, f_{,q}]\| \leq 1$ and substituting this into (23) we get

$$d(p,q) \geq |f_{,q}(p) - f_{,q}(q)| = d_\gamma(p,q).$$

This final equation along with (36) prove the result.

**Remarks:**
1. As a simple example consider $M = \mathbf{R}$ and $D = d/dx$. Then the condition $\|[D,f]\| \leq 1$ reads $\sup |df/dx| \leq 1$ and the supremum is saturated by the function $f(x) = x + \cos y$ which gives the usual distance.

2. Let us recall for brevity that the Lipschitz norm of some function f in C(M), where M is some closed Riemannian spin manifold, is given by the following formula:
$\|f\|_{Lip} = \|[d+d^*, \pi_1(f)]^2\|_{op} = (1/2) \|[[\Delta, \pi_2(f)], \pi_2(f)]\|_{op} = \|[D, \pi(f)]\|^2_{op}$ where the index "op" stands for the operator norm, d is the exterior derivative, d* is its (Hodge) adjoint (using the Riemannian metric), $\Delta = dd^* + d^*d$ denotes the Laplace operator, D denotes the Dirac operator and $\pi_1, \pi_2$ and $\pi$ denote the representations of the algebra C(M) onto the Hilbert spaces
$L^2 (M, \Lambda)$, $L^2(M)$ and $L^2 (M, S)$ of square integrable forms, functions and spinors respectively (where functions act multiplicatively in all cases).

3. A slightly different method to recover the geodesic distance between two points on the Riemannian spin manifold M would be to make use of the corresponding characters: Let x,y be to points on M; then from Gelfand's theorem these points correspond to two characters $\chi_x$ and $\chi_y$ respectively (namely $\chi_x$ and $\chi_y$ belong to the spectrum Spec(C(M)) of C(M)) defined as follows: $\chi_x(f) = f(x)$ and $\chi_y(f) = f(y)$ for every f in C(M). Then the geodesic distance d(x,y) between the two points x,y of $\Omega$ for the Riemannian metric g of $\Omega$ is given by the relation
$$d(x,y) = \sup \{ |\chi_x(f) - \chi_y(f)| ; f \in C(M), \|[D,f]\| \leq 1 \}$$ where the
norm $\|[D,f]\|$ is the operator norm in the Hilbert space H, namely
$$\|[D,f]\| = \min \{k \geq 0 : \|[D,f] v\| \leq k \|v\| \text{ for every element v of } \mathcal{H}\}.$$ An equivalent definition for the operator norm which has similarities with the original proof is that
$$\|[D,f]\| = \sup \{\|[D,f] v\| / \|v\|, v \text{ in H with } v \neq 0 \}.$$

4. The Dirac propagator (equations 17, 18) written in 4-dim Minkowski space in section 4 gives an opportunity to measure distances using the argument (x-y) which is the space distance separating two points. The fundamental Lemma proved in section 5 essentially declares that

$$ds = D^{-1},$$

namely length can be measured in the space of the corresponding quantum states by using the Dirac propagator and this is applied to curved manifolds as well as to noncommutative spaces which are not necessarily manifolds. In relativistic quantum field theory the propagator is the Green function of the corresponding wave operator of a particle and according to the Feynman path integral approach it is the basic ingredient in the expression that calculates the probability amplitude for a particle (in our case for a fermion, see [20] for example) to travel from one place to another in a given time (position space) or to travel with certain energy and momentum (momentum space).

The advantages of this approach for measuring distance and length are the following: This formalism is generally covariant and hence it might be proved useful in the future to measure time apart from spatial distance. GR is a generally covariant theory yet this is not currently the case for quantum field theory: Since there is no quantum theory of gravity yet there is a clear distinction between space and time coordinates (this is due to the existence of the arrow of time, the third thermodynamic law etc). In addition since this formalism involves the Dirac operator it enables one to use any fermion (although electrons are perhaps the most practically convenient fermions). It may be the case that spacetime is not a manifold after all but some noncommutative space; this formalism could also be applied in this case to make use of the full force of noncommutative geometry (see [18], [32] and [33]). Finally our approach has its roots in modern quantum physics (quantum field theory) which has taught us that the true invariant structure of nature (at least at small scales) is the spectra of the momentum and energy operators.